\documentclass[sigconf]{acmart}
\AtBeginDocument{%
  \providecommand\BibTeX{{%
    \normalfont B\kern-0.5em{\scshape i\kern-0.25em b}\kern-0.8em\TeX}}}

\copyrightyear{2023}
\acmYear{2023}

\usepackage{booktabs}
\usepackage{tcolorbox}
\usepackage{multirow}
\usepackage{bbding}
\usepackage{utfsym}
\usepackage{ulem}
\usepackage{xcolor}
\usepackage{url}
\usepackage{multirow}
\usepackage{tcolorbox}
\usepackage{graphicx}
\usepackage{enumitem}
\usepackage{xspace}
\usepackage{pifont}
\usepackage{wasysym}

\begin{document}

\title{Structured Chain-of-Thought Prompting for Code Generation}

\author{Jia Li \male}
\email{lijia@stu.pku.edu.cn}
\affiliation{%
  \institution{Peking University}
  \city{Beijing}
  \country{China}
}

\author{Ge Li}
\affiliation{%
  \institution{Peking University}
  \city{Beijing}
  \country{China}}
\email{lige@pku.edu.cn}

\author{Yongmin Li}
\affiliation{%
  \institution{Peking University}
  \city{Beijing}
  \country{China}}
\email{liyongmin@pku.edu.cn}

\author{Zhi Jin}
\affiliation{%
  \institution{Peking University}
  \city{Beijing}
  \country{China}}
\email{zhijin@pku.edu.cn}

\begin{abstract}
Large Language Models (LLMs) (\eg ChatGPT) have shown impressive performance in code generation. LLMs take prompts as inputs, and Chain-of-Thought (CoT) prompting is the state-of-the-art prompting technique. CoT prompting asks LLMs first to generate CoTs (\ie intermediate natural language reasoning steps) and then output the code. However, CoT prompting is designed for natural language generation and has low accuracy in code generation.

In this paper, we propose Structured CoTs (SCoTs) and present a novel prompting technique for code generation, named \method. Our motivation is source code contains rich structural information and any code can be composed of three program structures (\ie sequence, branch, and loop structures) \cite{Basic_structure}. 
Intuitively, structured intermediate reasoning steps make for structured source code.
Thus, we ask LLMs to use program structures to build CoTs, obtaining SCoTs. Then, LLMs generate the final code based on SCoTs. Compared to CoT prompting, \method explicitly 
constraints LLMs to think about how to solve requirements from the view of source code and further the performance of LLMs in code generation. We apply \method to two LLMs (\ie ChatGPT and Codex) and evaluate it on three benchmarks (\ie HumanEval, MBPP, and MBCPP). (1) \textbf{\method outperforms the state-of-the-art baseline - CoT prompting by up to 13.79\% in Pass@1.} (2) Human evaluation shows human developers prefer programs from \method. (3) \method is robust to examples and achieves substantial improvements.
\end{abstract}



\def\eg{\textit{e.g.,} }
\def\ie{\textit{i.e.,} }
\def\method{SCoT prompting\xspace}
\def\cot{CoT prompting\xspace}


\maketitle

\section{Introduction}
\label{sec:intro}

Code generation aims to automatically generate a program that satisfies a given natural language requirement \cite{SkCoder,CodeEditor,Self-Edit}.
Large Language Models (LLMs) have recently shown impressive performance in code generation, such as ChatGPT \cite{ChatGPT}, and CodeGen \cite{CodeGen}. 
During the inference, LLMs take a prompt as input that consists of several examples (\eg $<$requirement, code$>$ pairs) and a new requirement. LLMs learn code generation from examples and analogously generate a new program.
The performance of LLMs heavily relies on the prompt \cite{Calibrate_before_use}. Nowadays, how to make an effective prompt (\ie \textit{Prompting technique}) for code generation is still an open question.

\begin{figure}[t]
\centering
\includegraphics[width=\linewidth]{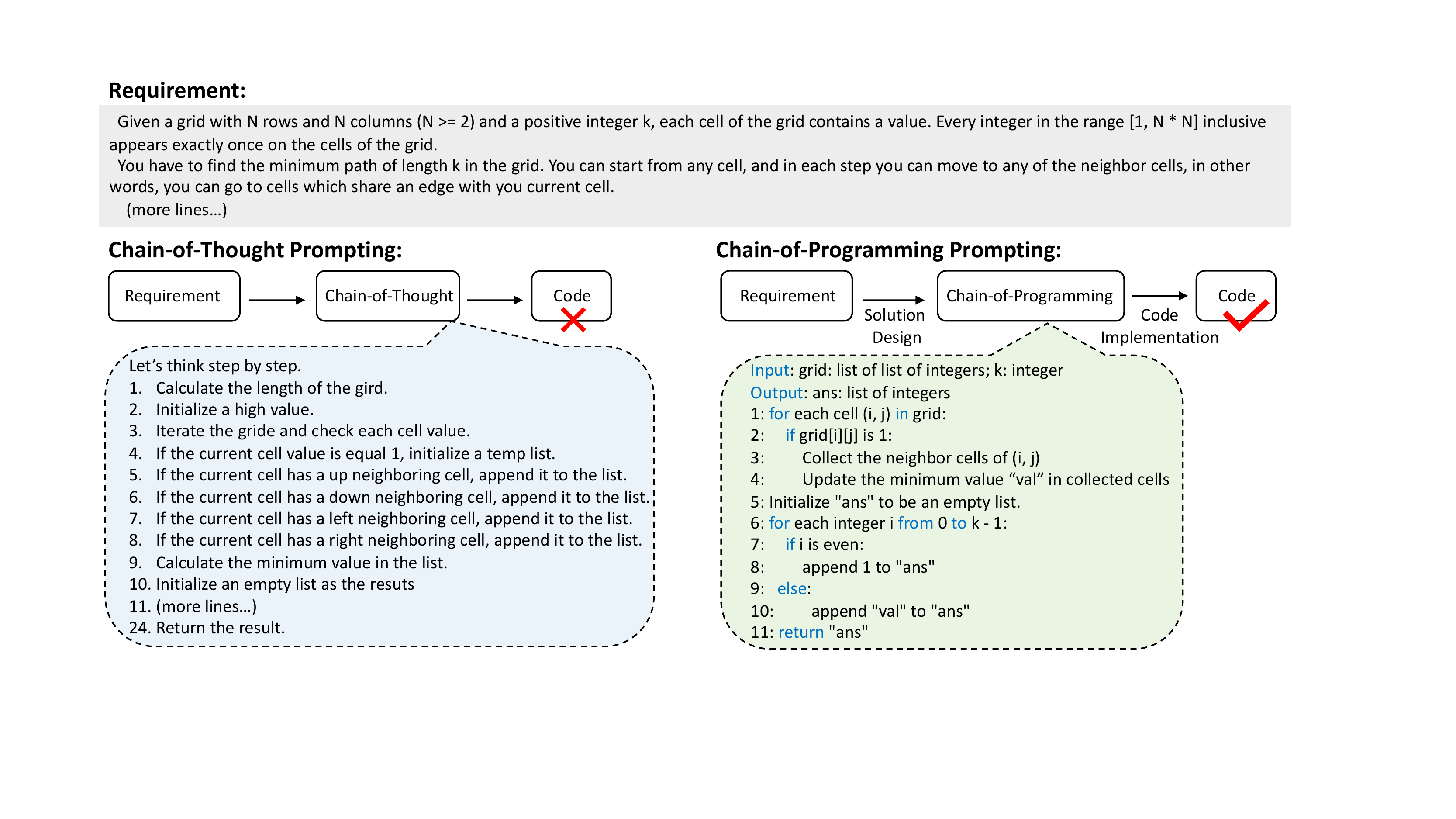}
\caption{The comparison of a Chain-of-Thoughts (CoT) and our Structured Chain-of-Thought (SCoT).}
\label{fig:intro_example}
\end{figure}

Chain-of-Thought (CoT) prompting \cite{chain-of-thought} is the state-of-the-art (SOTA) prompting technique. CoT Prompting asks LLMs first to generate a CoT and then output the code. A CoT is several intermediate natural language reasoning steps that describe how to write code step by step. Figure \ref{fig:intro_example} (a) shows a CoT on code generation. However, CoT prompting brings slight improvements in code generation. For example, it only improves ChatGPT by 0.82 points in Pass@1 upon a real-world benchmark \cite{Codex}.

\textbf{In this paper, we propose a Structured CoT for code generation.} Our motivation is that code generation aims to convert a natural language requirement to source code. Different from natural languages, source code contains rich structural information \cite{ASTNN,Modular_network,TPTrans}. For example, source code contains three basic structures \cite{Basic_structure}, including sequence, branch, and loop structures.
Intuitively, intermediate reasoning steps leading to the structured code should also be structured.
Consider a human developer's thought process when solving a requirement (\eg \texttt{find the maximum number in a list}). 
It is typical to come up with a solving process with program structures: \textit{``Initialize a result with -inf; for each number in the list; if the number is greater than result: Update result with the number ...''}.
Our idea is to enable LLMs to generate similar structured CoTs - a coherent series of intermediate reasoning steps constructed by program structures.
Besides, LLMs' training data contains lots of code data, so they have the ability to generate program structures. 
However, standard CoT ignores the program structures and has low accuracy in code generation.
Thus, it is necessary to design a structured CoT to unlock the reasoning ability of LLMs in code generation.

Figure \ref{fig:intro_example} (b) shows a SCoT. 
The design of our SCoT has two inspirations. First, existing work \cite{Basic_structure} proved that any simple or complex program can be composed of three basic structures, \ie sequence structure, branch structure, and loop structure. Thus, we introduce three basic structures and constrain LLMs to use them to generate CoTs. As shown in Figure \ref{fig:intro_example} (b), the SCoT uses a loop structure to clearly describe an iteration in line 2. While in the CoT, the scopes of two iterations in lines 2 and 4 are ambiguous. It shows the superiority of SCoT in code generation.
Besides, every program contains a required input-output structure, which includes the input-output parameters and their types (\eg \texttt{Input: array: list[list]; Output: result} in Figure \ref{fig:intro_example} (b)). By generating the input-output structure, LLMs are asked to analyze requirements and determine the entry and exit of the code, which benefits the following solving process.

\textbf{Based on the SCoT, we present a new prompting technique named \method.} It asks LLMs first to generate a SCoT using program structures and then implement the code. Compared to \cot, \method explicitly introduces program structures into intermediate reasoning steps and constraints LLMs to think about how to solve requirements from the view of programming languages. It further unlocks the reasoning ability of LLMs in code generation, thus achieving higher accuracy.

We apply \method to two popular LLMs (\ie ChatGPT \cite{ChatGPT} and Codex \cite{Codex}) and evaluate it on three representative benchmarks (\ie HumanEval \cite{Codex}, MBPP \cite{MBPP}, and MBCPP \cite{MBXP}). We use unit tests to measure the correctness of generated programs and report the Pass@$k$ ($k \in [1,3,5]$) \cite{Codex}. Based on experimental results, we obtain four findings. 
(1) \method significantly improves the accuracy of LLMs on code generation. \textbf{Compared to the SOTA baseline - Chain-of-Thought prompting, in terms of Pass@1, \method outperforms it by up to 13.79\% in HumanEval, 12.31\% in MBPP, and 6.63\% in MBCPP.}
(2) Human evaluation shows that human developers prefer programs generated by \method.
(3) \method is effective for different LLMs and different programming languages. In terms of Pass@1, it improves ChatGPT by up to 13.79\% and Codex by up to 13.77\%. Besides, \method is language-agnostic and effective in multiple languages (\eg Python and C++).
(4) We explore the robustness of \method to examples. Results show that \method does not depend on specific examples or writing styles.


We summarize our contributions in this paper as follows.
\begin{itemize}[leftmargin=*]
    \item We propose a Structured Chain-of-Thought (SCoT), which utilizes program structures to build the intermediate reasoning steps.
    \item We propose a novel prompting technique for code generation, named SCoT Prompting. It prompts large language models first to generate a SCoT and then implement the code.
    \item We conduct extensive experiments on three benchmarks. Qualitative and quantitative experiments show that \method significantly outperforms SOTA baselines (\eg Chain-of-Thought prompting).
    \item We discuss the contributions of different program structures and the robustness of \method.
\end{itemize}

\textbf{Data Availability.}
We open source our replication package \cite{anonymity2023}, including the datasets and the source code of \method, to facilitate other researchers and practitioners to repeat our work and verify their studies.

\section{Methodology}
\label{sec:approach}

In this section, we propose a Structured Chain-of-Thought (SCoT). A SCoT denotes several intermediate reasoning steps constructed by program structures. Then, we present a novel prompting technique for code generation named \method. \method asks LLMs first to generate a SCoT and then output the final code.
In the subsections, we first describe the design of our SCoT and further show the details of \method.

\subsection{Structured Chain-of-Thought}
\label{sec:approach:SCoT}


Standard Chain-of-Thought (CoT) is several intermediate natural language reasoning steps that lead to the final answer \cite{chain-of-thought}. The CoT is initially designed for natural language generation (\eg commonsense reasoning \cite{CommonsenseQA}). Thus, the CoT only uses natural languages to sequentially describe how to solve a problem step by step. Figure \ref{fig:intro_example} (a) shows a CoT on code generation. A limitation is that CoT brings slight improvements in code generation. For example, adding the CoT only improves ChatGPT by 0.82 points in Pass@1 upon a real-world benchmark - HumanEval \cite{Codex}.

In this paper, we propose a Structured CoT. Our motivation is that, unlike natural language generation, the goal of code generation is highly structured code. Source code solves a problem through special structures, including sequence structures, branch structures, and loop structures. 
For example, given a requirement - \texttt{reading text from a given file}, imagine a human developer's thought process.
The developer will use program structures to design an initial idea: \textit{``if the given file exists: read text from the file; else: raise an error;''}. The program structures clearly show the solving process and benefit the following code implementation. Thus, intermediate reasoning steps leading to the code should also be structured.

\begin{figure}[t]
\centering
\includegraphics[width=\linewidth]{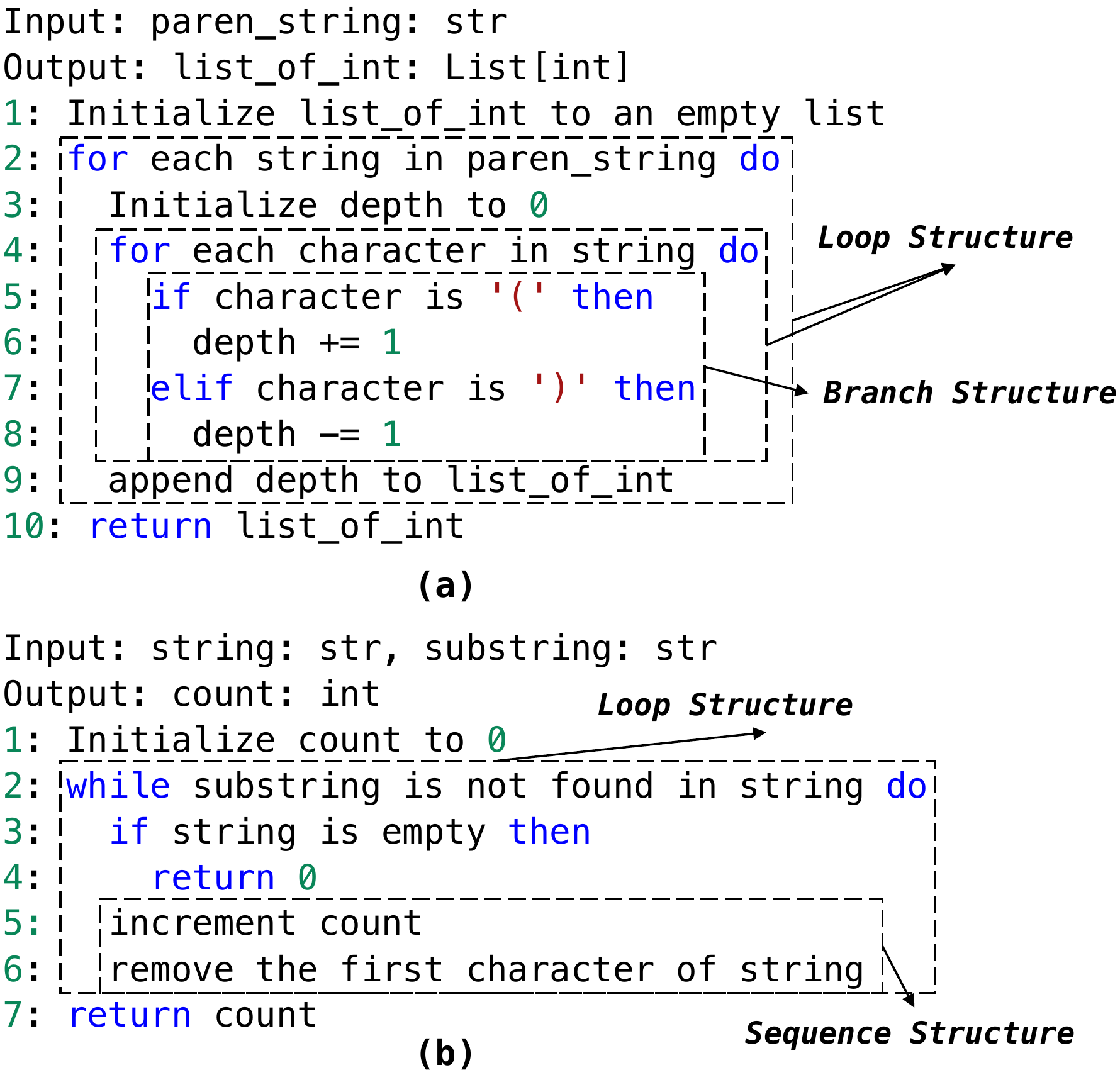}
\caption{Examples of SCoT in code generation.}
\label{fig:SCoT_ex}
\end{figure}

Figure \ref{fig:SCoT_ex} shows some examples of SCoT. Compared to the CoT, our SCoT explicitly introduces program structures. 
Existing work \cite{Basic_structure} proved that any simple
or complex program can be composed of three basic structures,
\ie sequence structure, branch structure, and loop structure
Thus, we introduce three basic structures, whose details are shown as follows.
\begin{itemize}[leftmargin=*]
    \item \textbf{Sequence Structure.} The intermediate steps are sequentially placed and all steps are at the same level.
    \item \textbf{Branch Structure.} It starts with a condition and places different intermediate steps for different results of the condition. In this paper, branch structures contain three formats, \ie \texttt{if ...}, \texttt{if ... else}, and \texttt{if ... elif ... else}.
    \item \textbf{Loop Structure.} A set of intermediate steps are repeatedly conducted until given conditions are not met. In this paper, loop structures contain two basic formats, including the \texttt{for loop} and the \texttt{while loop}.
\end{itemize}

We allow the nesting between different program structures. It allows LLMs to design more complex SCoT for some difficult requirements. As shown in Figure \ref{fig:SCoT_ex}, the SCoT flexibly uses various program structures to build a solving process.

Besides three basic structures, we add the input-output structure, which contains input-output parameters and their types. Our motivation is that an input-output structure is required for a program, which indicates the entry and exit. 
Generating the input-output structure is beneficial to clarify requirements and generate the following solving process.

\subsection{\method}
\label{sec:approach:prompting}

Based on the SCoT, we propose a new prompting technique for code generation, named \method. It asks LLMs first to generate a SCoT and then output the final code.

\begin{figure}[t]
\centering
\includegraphics[width=\linewidth]{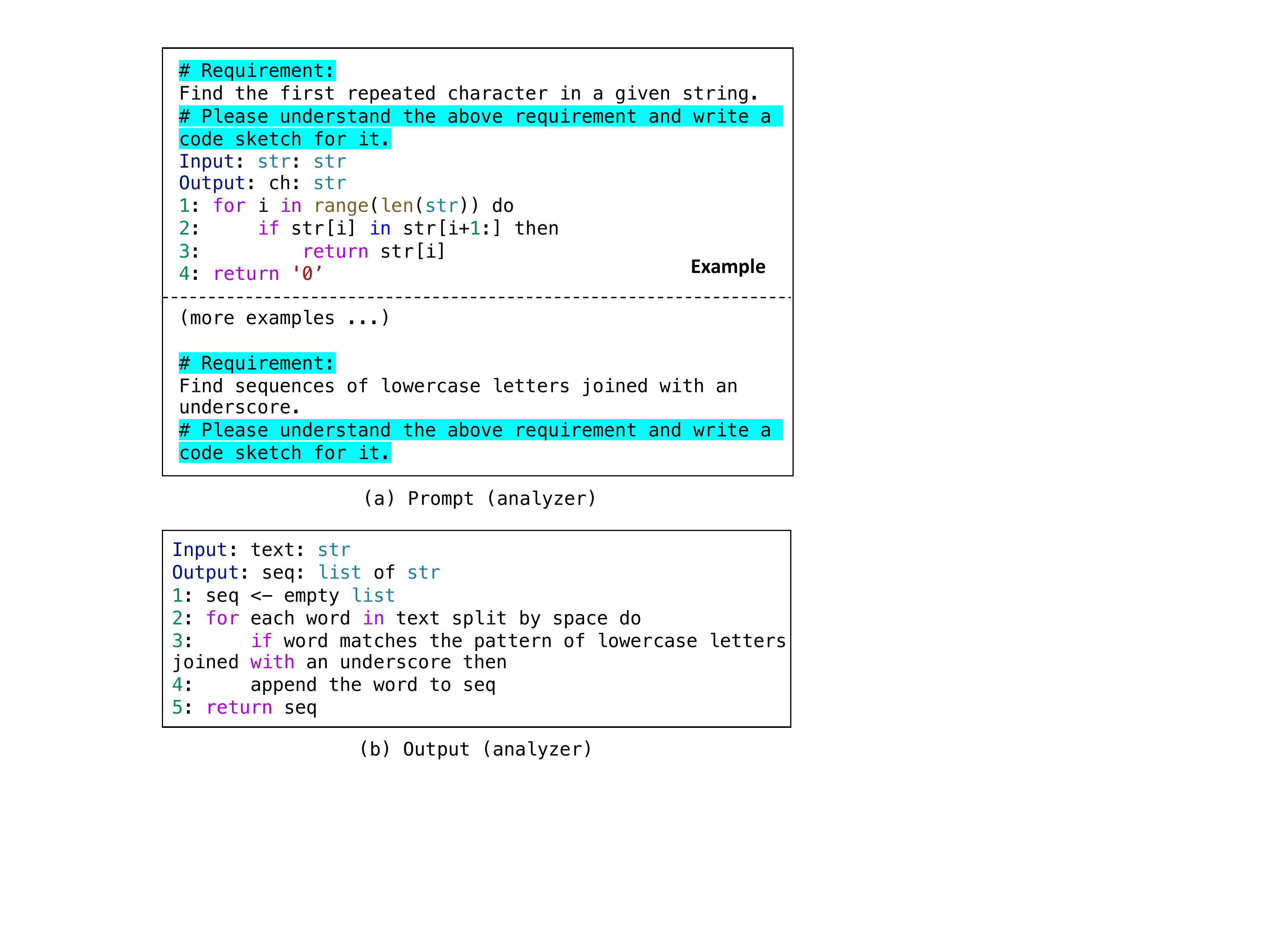}
\caption{A prompt for generating a SCoT.}
\label{fig:prompt4SCoT}
\end{figure}

To implement \method, we design two special prompts. 
The first prompt is used to generate a SCoT, and an example of the prompt is shown in Figure \ref{fig:prompt4SCoT}.
The prompt starts with several examples (\ie $<$requirement, SCoTs$>$). These examples cover three basic program structures and the input-output structure. Next, the 
\textit{italic sentences} are instructions for LLMs, which indicate the goal of LLMs and related constraints. Finally, the prompt ends with a new requirement and is fed into LLMs. We expect LLMs to learn from examples and generate a SCoT for the new requirement.

\begin{figure}[t]
\centering
\includegraphics[width=\linewidth]{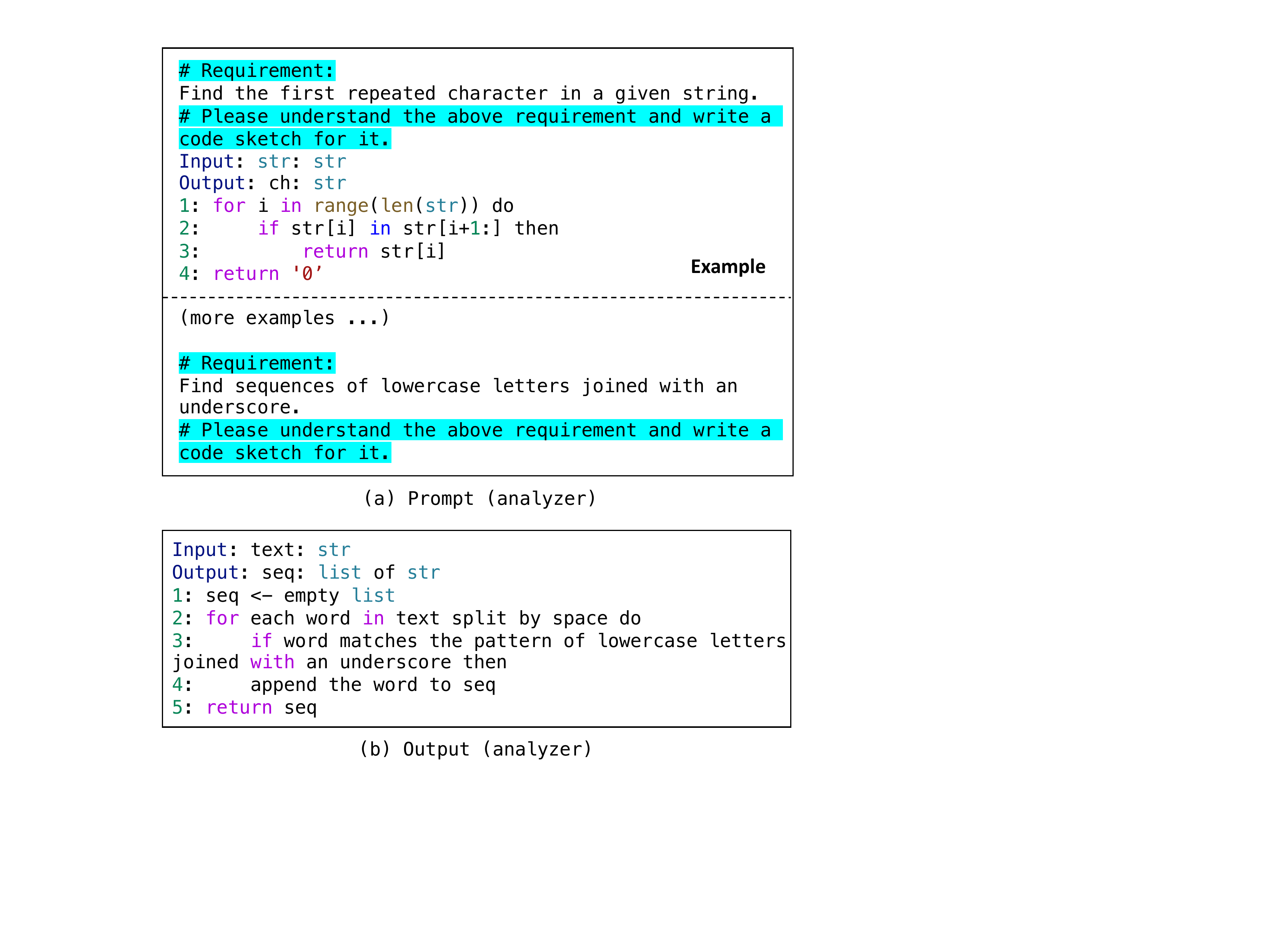}
\caption{A prompt for generating the code.}
\label{fig:prompt4code}
\end{figure}

After generating a SCoT, we design the second prompt for generating the final code. An example of the prompt is shown in Figure \ref{fig:prompt4code}. The prompt starts with several examples (\ie $<$requirement, SCoT, code$>$). The \textit{italic sentences} are instructions. We consider the SCoT as a soft template and ask LLMs to implement a program. Finally, the prompt ends with a new requirement and its SCoT, and is input into LLMs. By learning from examples, LLMs generate a new program based on the requirement and SCoT.

Related work \cite{ErrorAccum} has found that generative models may be negatively affected by error accumulation. Similarly, in \method, the generated SCoT may contain noises (\eg errors or missing steps). These noises will further negatively affect code implementation. In this paper, we utilize two approaches to alleviating error accumulation. First, as shown in Figure \ref{fig:prompt4code}, we ask LLMs to double-check the SCoT and fix possible noises. It allows LLMs to adaptively refer to the SCoT and filter out noises. Second, \method utilizes a two-step generation pipeline. It provides a window of opportunity to debug where the SCoT goes wrong. In practice, human developers can first check the generated SCoT and fix possible errors. Then, the SCoT is used to generate code.

\subsection{Implementation Details}
\label{sec:approach:implementation_detail}

\method is a prompting technique for code generation, which does not rely on specific LLMs. In this paper, we consider ChatGPT as the default LLM. 
We select a few (\eg three) $<$requirement, code$>$ pairs from real-world benchmarks (\ie training data) as example seeds. Then, we manually write the SCoT for seeds and obtain examples - $<$requirement, SCoT, code$>$ triples, which are used to make prompts in Figure \ref{fig:prompt4SCoT} and \ref{fig:prompt4code}.
A prompt contains three examples by default. The examples and prompt templates are available in our replication
package. In the future, users can flexibly apply our approach to more powerful LLMs in a plug-and-play fashion.

\section{Study Design}
\label{sec:study_design}

To assess \method, we conduct a large-scale study to answer four research questions. In this section, we present the details of our study, including datasets, evaluation metrics, comparison baselines, and implementation details.

\subsection{Research Questions}
\label{sec:study_design:RQ}

Our study aims to answer the following research questions (RQ).

\textbf{RQ1: How does \method perform in terms of accuracy compared to baselines?}
This RQ aims to verify that \method has a higher accuracy than existing prompting techniques on code generation.
We apply three existing prompting techniques and \method to two LLMs. Then, we use unit tests to measure the correctness of programs generated by different approaches and report the Pass@k.

\textbf{RQ2: Do developers prefer programs generated by \method?}
The ultimate goal of code generation is to assist human developers in writing code. In this RQ, we hire 10 developers (including industry employees and academic researchers) to manually review the programs generated by \method and baselines. We measure the quality of programs in three aspects, including correctness, code smell, and maintainability.


\textbf{RQ3: Is \method robust to examples?}
Prompting techniques may be sensitive to example \cite{Calibrate_before_use}. In this RQ, we measure the robustness of \method to examples. Specifically, we measure the performance of \method with different example seeds and different example writing styles.

\textbf{RQ4: What are the contributions of different program structures in \method?} As stated in Section \ref{sec:approach:SCoT}, \method introduces three basic structures and the input-output structure. This RQ is designed to analyze the contributions of different structures. We select an LLM as the base model. Then, we individually remove a program structure and report the fluctuations in performance.

\begin{table}[t]
\centering
\caption{Statistics of the datasets in our experiments.}
\begin{tabular}{lccc}
\toprule
Statistics                 & HumanEval & MBPP & MBCPP  \\
\midrule
Language                   & Python & Python & C++ \\
\midrule
\# Train                   & --  & 474   & 413  \\
\# Test                    & 164  & 500   & 435  \\
\midrule
Avg. tests per sample      & 7.7       & 3  & 3 \\
\bottomrule
\end{tabular}
\label{tab:dataset}
\end{table}

\subsection{Benchmarks}
\label{sec:study_design:data}

Following previous studies \cite{Codex,CodeGen,CodeGeeX,CodeT}, we conduct experiments on three representative code generation benchmarks, including the HumanEval in Python, MBPP in Python, and MBCPP in C++. 
The details of the benchmarks are described as follows.
\begin{itemize}[leftmargin=*]
    \item \textbf{HumanEval} \cite{Codex} is a Python function-level code generation benchmark, which contains 164 hand-written programming problems. Each programming problem consists of an English requirement, a function signature, and several test cases, with an average of 7.7 test cases per problem.
    \item \textbf{MBPP} \cite{MBPP} is a Python function-level code generation benchmark. It contains 974 programming problems that involve simple numeric manipulations or basic usage of standard libraries. Each problem contains an English requirement, a function signature, and three manually written test cases for checking functions.
    \item \textbf{MBCPP} \cite{MBXP} is a C++ function-level code generation benchmark. It consists of 848 programming problems that are collected by crowd-sourcing. Each problem contains an English description, a function signature, and three test cases for checking the correctness of functions.
\end{itemize}

We follow the original splits of three datasets. The statistics of the benchmarks are shown in Table \ref{tab:dataset}. We randomly pick several samples from training data to make examples in prompts (Section \ref{sec:approach:implementation_detail}). Then, we measure the performance of different approaches on test data. Because HumanEval does not contain train data, we reuse examples from MBPP in HumanEval.

\subsection{Evaluation Metrics}
\label{sec:study_design:metric}

Following previous code generation studies \cite{Codex, CodeGen, CodeGeeX, CodeT}, we use Pass@$k$ as our evaluation metrics. Specifically, given a requirement, a code generation model is allowed to generate $k$ programs. The requirement is solved if any generated programs pass all test cases. We compute the percentage of solved requirements in total requirements as Pass@$k$. For Pass@$k$, a higher value is better. In our experiments, $k$ is set to 1, 3, and 5, because we think that developers mainly use Top-5 outputs in real-world scenarios.

Previous work \cite{Codex, MBXP, CodeT} found that standard Pass@$k$ has high variance and proposed an unbiased Pass@$k$. We follow previous work and employ the unbiased Pass@$k$.
Specifically, we generate $n \geq k$ programs per requirement (in this paper, we use $n=20$, $k \in [1, 3, 5]$), count the number of solved requirements $c$, and calculate the unbiased Pass@$k$:
\begin{equation}
\text{Pass} @ k:=\underset{\text { Problems }}{\mathbb{E}}\left[1-\frac{\left(\begin{array}{c}
n-c \\
k
\end{array}\right)}{\left(\begin{array}{l}
n \\
k
\end{array}\right)}\right]
\end{equation}

We also notice that previous code generation studies use text-similarity-based metrics (\eg BLEU \cite{BLEU}). These metrics are initially designed for natural language generation and are poor in measuring the correctness of programs \cite{Codex}. Thus, we omit these metrics in our experiments.

\subsection{Comparison Baselines}
\label{sec:study_design:baseline}

This paper proposes a new prompting technique for code generation. To assess the effectiveness of our approach, we select three mainstream prompting techniques as baselines.
\begin{itemize}[leftmargin=*]
    \item \textbf{Zero-shot prompting} \cite{Codex} directly feeds the requirement into LLMs without examples. Then, it extracts a generated program from LLMs' outputs.
    \item \textbf{Few-shot prompting} \cite{Codex} randomly selects several $<$ requirement, code$>$ pairs as examples. Given a requirement, it concatenates examples and the requirement together, making a prompt. Then, the prompt is fed into LLMs, and LLMs predict a new program.
    \item \textbf{Chain-of-Thought (CoT) prompting} \cite{chain-of-thought} is a variant of few-shot prompting. CoT prompting produces a special prompt consisting of $<$requirement, CoT, code$>$ triples as examples. A CoT is several intermediate natural language reasoning steps. 
\end{itemize}
To ensure the fairness of comparison, all baselines and \method 
have the same number of examples (\ie three examples) and example seeds.

The criteria for selecting baselines are three-fold. (1) \method is a prompting technique for code generation. Thus, we directly compare it to existing prompting techniques for code generation. We also notice some emerging prompting techniques in other fields, such as Self-Consistency \cite{Self-Consistency} and Least-to-Most \cite{Least-to-Most}. But these approaches are designed for specific tasks (\eg Arithmetic reasoning) and can not be directly applied to code generation. Thus, we omit them in this paper. (2) Our approach is to augment LLMs and can be flexibly applied to different LLMs. Thus, we do not directly compare LLMs to our approach. (3) We also omit some rank techniques for code generation \cite{CodeT}. They first use LLMs to generate many candidates and then leverage test cases or neural networks to rerank candidates. We think our work and these rank techniques are complementary. Users can use our approach to generate programs and then use post-processing techniques to select the final output. We further discuss the complementarity through experiments in Section \ref{sec:discussion:rank}.

\begin{table*}[t]
\caption{The Pass@k (\%) of \method and baselines on three code generation benchmarks. The numbers in {\color[HTML]{FE0000} red} denote \method's relative improvements compared to the SOTA baseline - \cot.}
\label{tab:RQ1}
\begin{tabular}{ccccccccccc}
\toprule
 &  & \multicolumn{3}{c}{HumanEval} & \multicolumn{3}{c}{MBPP} & \multicolumn{3}{c}{MBCPP} \\
\multirow{-2}{*}{Base Model} & \multirow{-2}{*}{Prompting Technique} & Pass@1 & Pass@3 & Pass@5 & Pass@1 & Pass@3 & Pass@5 & Pass@1 & Pass@3 & Pass@5 \\ \midrule
 & Zero-shot prompting & 49.73 & 66.07 & 71.54 & 37.07 & 43.54 & 48.58 & 47.53 & 60.09 & 64.22 \\
 & Few-shot prompting & 52.47 & 69.32 & 74.10 & 40.00 & 49.82 & 53.13 & 52.58 & 63.03 & 66.11 \\
 & CoT prompting & 53.29 & 69.76 & 75.52 & 41.83 & 51.04 & 54.57 & 53.51 & 63.84 & 67.03 \\
\multirow{-4}{*}{ChatGPT} & SCoT Prompting & \textbf{60.64} & \textbf{73.53} & \textbf{77.32} & \textbf{46.98} & \textbf{55.31} & \textbf{58.36} & \textbf{57.06} & \textbf{65.70} & \textbf{68.70} \\ \midrule
\multicolumn{2}{c}{Relative Improvement} & {\color[HTML]{FE0000} \textbf{13.79\%}} & {\color[HTML]{FE0000} \textbf{5.40\%}} & {\color[HTML]{FE0000} \textbf{2.38\%}} & {\color[HTML]{FE0000} \textbf{12.31\%}} & {\color[HTML]{FE0000} \textbf{8.37\%}} & {\color[HTML]{FE0000} \textbf{6.95\%}} & {\color[HTML]{FE0000} \textbf{6.63\%}} & {\color[HTML]{FE0000} \textbf{2.91\%}} & {\color[HTML]{FE0000} \textbf{2.49\%}} \\ \midrule
 & Zero-shot prompting & 40.20 & 61.78 & 68.11 & 27.07 & 43.81 & 47.93 & 40.25 & 54.17 & 60.65 \\
 & Few-shot prompting & 42.93 & 62.96 & 70.10 & 33.17 & 45.72 & 49.62 & 44.12 & 57.65 & 62.45 \\
 & CoT prompting & 43.79 & 63.41 & 71.56 & 35.66 & 46.57 & 50.11 & 45.79 & 58.92 & 62.56 \\
\multirow{-4}{*}{Codex} & SCoT Prompting & \textbf{49.82} & \textbf{66.56} & \textbf{75.14} & \textbf{38.29} & \textbf{50.74} & \textbf{53.16} & \textbf{48.34} & \textbf{60.77} & \textbf{64.19} \\ \midrule
\multicolumn{2}{c}{Relative Improvement} & {\color[HTML]{FE0000} \textbf{13.77\%}} & {\color[HTML]{FE0000} \textbf{4.97\%}} & {\color[HTML]{FE0000} \textbf{5.00\%}} & {\color[HTML]{FE0000} \textbf{7.38\%}} & {\color[HTML]{FE0000} \textbf{8.95\%}} & {\color[HTML]{FE0000} \textbf{6.09\%}} & {\color[HTML]{FE0000} \textbf{5.57\%}} & {\color[HTML]{FE0000} \textbf{3.14\%}} & {\color[HTML]{FE0000} \textbf{2.61\%}} \\
\bottomrule
\end{tabular}
\end{table*}

\subsection{Base Large Language Models}
\label{sec:study_design:base_model}

There are many available LLMs for source code. Our motivation is that existing LLMs can be divided into two categories: standard language models and instruction-tuned models. For each category, we pick a representative model as the base model.

(1) Standard language models are pre-trained on a large-scale corpus with the next-token prediction objective. They are mainly used to continually complete the given content, such as code completion. Thus, we pick the state-of-the-art completion model for code - Codex \cite{Codex} as a base model.

\textbf{Codex} \cite{Codex} is a powerful language model for code generation, which supports a commercial application - GitHub Copilot \cite{Copilot}. Codex's training data contains both natural language and billions of lines of code. We use OpenAI's APIs to access the latest version of Codex with 175 billion parameters, \ie code-davinci-002 \cite{Codex-API}.

(2) Instruction-tuned models refer to LLMs after instruction tuning. Instruction tuning trains LLMs to understand human users' instructions and perform tasks based on the instructions. We select the state-of-the-art instruction-tuned model - ChatGPT \cite{ChatGPT} as a base model.

\textbf{ChatGPT} \cite{ChatGPT} is the state-of-the-art LLM for code generation. ChatGPT is trained with extensive natural language text and code files. Then, it is trained with reinforcement learning and learns to follow human instructions. We use OpenAI's APIs to access the ChatGPT, \ie gpt-3.5-turbo-0301 \cite{ChatGPT}. 

Our approach does not rely on specific LLMs and can be applied to different LLMs in a plus-and-play fashion. In the future, we will explore its usage on more powerful LLMs.

\subsection{Sampling Settings}
\label{sec:study_design:sampling}

Following previous studies \cite{Codex,CodeGeeX,CodeGen}, we use nucleus sampling \cite{nucleus_sampling} to decode programs from LLMs. To ensure the fairness of experiments, all baselines and \method generate 20 programs per requirement. The details of sampling settings are shown as follows.

\textbf{Baselines.} The temperature is 0.8 and the top-$p$ is 0.95. For zero-shot and few-shot prompting, the maximum generated length is 300 tokens. The maximum generated length of CoT prompting is 600 tokens. Our motivation is that CoT prompting needs to generate intermediate reasoning steps and code. Thus, it requires a larger generation length.

\textbf{\method.} In the first step, we sample 20 individual SCoTs from LLMs per requirement. The temperature is 0.8 and the top-$p$ is 0.95. The maximum generated length is 300 tokens. Then, for each SCoT, we use LLMs to generate a corresponding program. The temperature is 0 and the maximum generated length is 300 tokens. Finally, we obtain 20 programs for each requirement. The total generation length of two steps is the same as CoT prompting.

\section{Results and Analysis}
\label{sec:results}

\subsection{RQ1: How does \method perform in terms of accuracy compared to baselines?}
\label{sec:results:RQ1}

In the first research question, we apply \method and baselines
to three benchmarks and use unit tests to measure the correctness of generated programs.

\textbf{Setup.} We apply baselines and \method to two LLMs (Section \ref{sec:study_design:base_model}). Then, we measure the performance of different approaches on three code generation benchmarks (Section \ref{sec:study_design:data}) using the Pass@k (Section \ref{sec:study_design:metric}).

\textbf{Results.} The Pass@$k$ ($k \in [1,3,5]$) of different approaches are shown in Table \ref{tab:RQ1}. The numbers in {\color[HTML]{FE0000} red} denote \method's relative improvements compared to the SOTA baseline - \cot.

\textbf{Analyses.} 
\uline{(1) \method achieves the best results among all baselines.} 
Table \ref{tab:RQ1} shows that \method can generate more correct programs than baselines on three benchmarks. Compared to the SOTA baseline - \cot, in terms of Pass@1, \method outperforms it by up to 13.79\% in HumanEval, 12.31\% in MBPP, and 6.63\% in MBCPP. Pass@1 is a strict metric and it is difficult to improve. The results show that \method can significantly improve the accuracy of LLMs on code generation and is more promising than existing prompting techniques.
\uline{(2) \method is effective in different LLMs and programming languages.} 
\method is effective in different LLMs. Compared to \cot, in terms of Pass@1, \method further improves ChatGPT by up to 13.79\% and Codex by up to 13.77\%. Besides, \method is language-agnostic and can be applied to different programming languages. As shown in Table \ref{tab:RQ1}, \method brings substantial improvements in Python (\ie HumanEval and MBPP) and C++ (\ie MBCPP).
\uline{(3) \method unlocks the reasoning ability of LLMs on code generation.}
LLMs can benefit from generating intermediate reasoning steps. The baseline - \cot utilizes natural language steps but only brings slight improvements. In terms of Pass@1, \cot improves few-shot prompting by up to 2\% in HumanEval, 7.51\% in MBPP, and 3.79\% in MBCPP.
In this paper, we introduce program structures into intermediate reasoning steps and propose a Structured Chain-of-Thought (SCoT). The SCoT constrains LLMs to use program structures to generate intermediate steps, moving in the direction of code.
In terms of Pass@1, \method improves few-shot prompting by up to 16.05\% in HumanEval, 17.45\% in MBPP, and 9.56\% in MBCPP.
The improvements show that \method further unlocks the reasoning ability of LLMs in code generation. 

\begin{tcolorbox}[size=title]
\textbf{Answer to RQ1:} \method achieves higher accuracy than baselines on three benchmarks. In terms of Pass@1, \method outperforms the SOTA baseline by up to 13.79\% in HumanEval, 12.31\% in MBPP, and 6.63\% in MBCPP. The significant improvements show the effectiveness of \method in code generation.
\end{tcolorbox}

\subsection{RQ2: Do developers prefer programs generated by \method?}
\label{sec:results:RQ2}

The ultimate goal of code generation is to assist developers in writing programs. In this RQ, we hire 10 developers (including industry employees and academic researchers) to
manually review the programs generated by \method and
baselines.

\begin{table}[t]
\centering
\caption{The results of human evaluation in three aspects. The numbers in {\color[HTML]{FE0000} red} denote \method's relative improvements compared to the SOTA baseline - \cot. All the $p$-values are substantially smaller than 0.05.}
\resizebox{\linewidth}{!}{
\begin{tabular}{lccc}
\toprule
Approach   & Correctness & Code Smell & Maintainability \\ \midrule
Zero-shot prompting   & 1.012  & 1.523        & 1.372       \\
Few-shot prompting    & 1.119  & 1.653        & 1.552       \\
CoT prompting & 1.225  & 1.689        & 1.616       \\
\method    & \textbf{1.412} & \textbf{1.869}       & \textbf{1.873}    \\ \midrule
{Relative Improvement}     & {\color[HTML]{FE0000} \textbf{15.27\%}}      & {\color[HTML]{FE0000} \textbf{10.66\%}} & {\color[HTML]{FE0000} \textbf{15.90\%}} \\ \bottomrule
\end{tabular}}
\label{tab:RQ2}
\end{table}

\textbf{Setup.}
To ensure the fairness of evaluation, we follow settings of human evaluation in previous studies \cite{AixBench,SkCoder}. We have carefully checked the evaluation settings and think our settings are reliable. Specifically, we manually evaluate generated programs in the following aspects:
\begin{itemize}[leftmargin=*]
    \item \textbf{Correctness (whether the program satisfies the requirement).} 0 point: the program is totally inconsistent with the requirement. 1 point: the program is implemented, but misses some details. 2 points: the program is correctly implemented.
    \item \textbf{Code Smell (whether the program contains bad code smell).} 0 point: There is a serious code smell. 1 point: some details are not in place. There is code smell of low severity. 2 points: the details are in place. No obviously better code in terms of performance exists. If possible, resources are released accordingly. No obvious code smell.
    \item \textbf{Maintainability (whether the implementation is standardized and has good readability).} 0 point: the program does not follow a consistent specification, or there are many meaningless names in variable naming, or there are certain repetitions and redundant code. 1 point: the program implementation meets certain specifications. But some variable names can be further refined. 2 points: the program implementation is relatively standardized. The variable naming is basically semantically straightforward, and the readability is good.
\end{itemize}

We explain the above aspects to evaluators through some examples. We also discuss with evaluators and set the score of each aspect to an integer, ranging from 0 to 2 (from bad to good). For \method and baselines, we select a fixed LLM as the base model (\ie ChatGPT) and collect 200 generated programs per approach. Finally, we obtain 800 programs for evaluation. 
We invite 10 developers with 3-5 years of development experience to evaluate the programs in the form of a questionnaire.
The evaluators include industry employees and academic researchers that are not co-authors of this paper.
The 800 programs are divided into 5 groups, with each questionnaire containing one group. The programs are randomly shuffled and anonymously reviewed by evaluators.
Each group is evaluated by two evaluators, and the final score is the average of two evaluators' scores. Evaluators are allowed to search the Internet for unfamiliar concepts.

\begin{figure}[t]
\centering
\includegraphics[width=\linewidth]{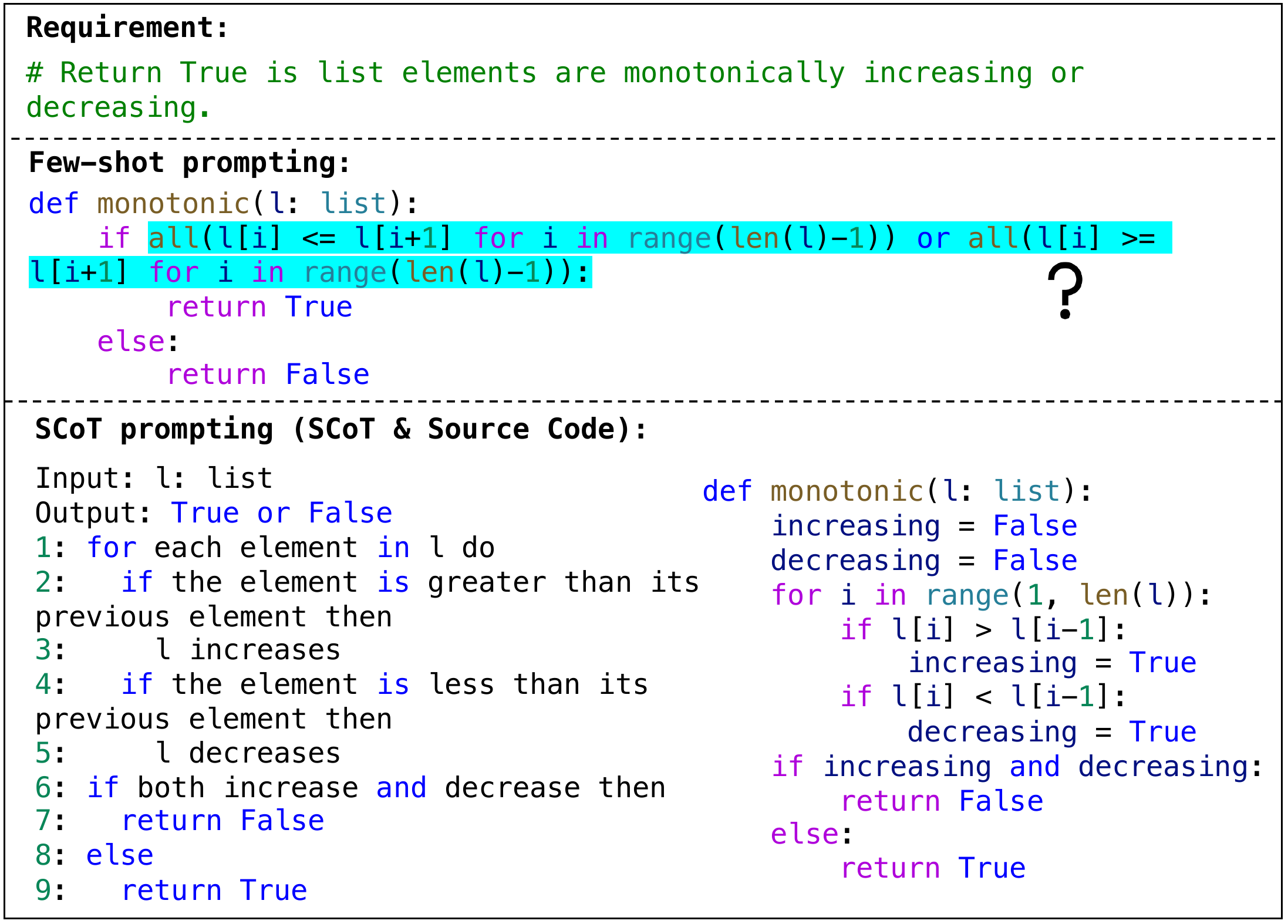}
\caption{Two programs generated by few-shot prompting and \method, respectively.}
\label{fig:maintain_ex}
\end{figure}

\begin{table*}[t]
\caption{The results of ablation study. The base model is ChatGPT.}
\label{tab:RQ4}
\begin{tabular}{lccccccccc}
\toprule
\multicolumn{1}{c}{\multirow{2}{*}{Prompting Technique}} & \multicolumn{3}{c}{HumanEval} & \multicolumn{3}{c}{MBPP} & \multicolumn{3}{c}{MBCPP} \\
\multicolumn{1}{c}{} & Pass@1 & Pass@3 & Pass@5 & Pass@1 & Pass@3 & Pass@5 & Pass@1 & Pass@3 & Pass@5 \\ \midrule
CoT prompting & 53.29 & 69.76 & 75.52 & 41.83 & 51.04 & 54.57 & 53.51 & 63.84 & 67.03 \\
SCoT prompting & \textbf{60.64} & \textbf{73.53} & \textbf{77.32} & \textbf{46.98} & \textbf{55.31} & \textbf{58.36} & \textbf{57.06} & \textbf{65.70} & \textbf{68.70} \\
\quad w/o Basic structures & 55.67 & 70.94 & 76.13 & 43.36 & 53.64 & 56.57 & 54.79 & 64.32 & 67.77 \\
\quad w/o IO structure & 59.65 & 72.79 & 77.12 & 46.13 & 54.76 & 57.88 & 56.61 & 65.01 & 68.42 \\
\bottomrule
\end{tabular}
\end{table*}

\textbf{Results.}
The results of the human evaluation are shown in Table \ref{tab:RQ2}. The numbers in {\color[HTML]{FE0000} red} denote \method's relative improvements compared to the SOTA baseline - \cot. All the $p$-values are substantially smaller than 0.05.

\textbf{Analyses.}
\uline{\method achieves the highest scores in all three aspects among all baselines.}
Specifically, \method outperforms the SOTA baseline - \cot by 15.27\% in correctness, 10.66\% in code smell, and 15.90\% in maintainability. 

We attribute the improvements to our proposed SCoT. The SCoT constrains LLMs to use program structures to generate intermediate reasoning steps. It allows LLMs to explore diverse solutions with three basic structures, improving the correctness of the code. Then, based on the SCoT, LLMs focus on implementing a program in a standardized way. Thus, the generated programs contain fewer code smells than ones from baselines.

Figure \ref{fig:maintain_ex} shows two programs generated by \method and few-shot prompting, respectively. Both programs pass unit tests. But the program from few-shot prompting contains a very complex statement highlighted in Figure \ref{fig:maintain_ex}). Developers have to spend lots of effort to understand and maintain this program.
In contrast, the program from \method has good readability, and the SCoT clearly explains the behavior of the code. Developers can further use the SCoT as comments of the program for future maintenance.

\begin{tcolorbox}[size=title]
\textbf{Answer to RQ2:} Human developers prefer programs generated by \method. Specifically, \method outperforms the SOTA baseline by 19.93\% in correctness, 11.25\% in code smell, and 16.17\% in maintainability. A case study also shows the program from \method is easy to read and maintain.
\end{tcolorbox}

\begin{table}[t]
\caption{The Pass@k of \cot and \method with different example seeds.}
\label{tab:RQ3_example}
\vspace{-0.2cm}
\resizebox{\linewidth}{!}{
\begin{tabular}{lcccccc}
\toprule
\multicolumn{1}{c}{\multirow{2}{*}{Seed}} & \multicolumn{3}{c}{CoT prompting} & \multicolumn{3}{c}{SCoT prompting} \\ 
\multicolumn{1}{c}{} & Pass@1 & Pass@3 & Pass@5 & Pass@1 & Pass@3 & Pass@5 \\ \midrule
Seed A & 53.29 & 69.76 & 75.52 & \textbf{60.64} & \textbf{73.53} & \textbf{77.32} \\
Seed B & 52.81 & 68.97 & 74.55 & \textbf{60.27} & \textbf{73.11} & \textbf{77.16} \\
Seed C & 51.36 & 67.44 & 73.62 & \textbf{59.36} & \textbf{72.88} & \textbf{76.79} \\
\bottomrule
\end{tabular}}
\end{table}

\begin{table}[t]
\caption{The Pass@k of \cot and \method with different annotators.}
\label{tab:RQ3_annotator}
\vspace{-0.2cm}
\resizebox{\linewidth}{!}{
\begin{tabular}{lcccccc}
\toprule
\multicolumn{1}{c}{\multirow{2}{*}{Annotator}} & \multicolumn{3}{c}{CoT prompting} & \multicolumn{3}{c}{SCoT prompting} \\ 
\multicolumn{1}{c}{} & Pass@1 & Pass@3 & Pass@5 & Pass@1 & Pass@3 & Pass@5 \\ \midrule
Annotator A & 53.29 & 69.76 & 75.52 & \textbf{60.64} & \textbf{73.53} & \textbf{77.32} \\
Annotator B & 51.43 & 67.92 & 73.44 & \textbf{59.48} & \textbf{72.16} & \textbf{76.44} \\
Annotator C & 52.18 & 68.45 & 74.71 & \textbf{60.02} & \textbf{73.15} & \textbf{77.24} \\
\bottomrule
\end{tabular}}
\end{table}

\subsection{RQ3: Is \method robust to examples?}
\label{sec:results:RQ3}

As stated in Section \ref{sec:approach:implementation_detail}, \method requires manually written examples to make prompts. In practice, people may write different examples, which makes the performance of \method varies. Thus, in this RQ, we explore the robustness of \method to examples.

\textbf{Setup.} 
As stated in Section \ref{sec:approach:implementation_detail}, we select some $<$requirement, code$>$ pairs as example seeds and manually write SCoTs for them, obtaining examples in prompts. In this RQ, we measure the robustness of \method to examples in two aspects, \ie seed selection and writing style.

\textbf{(1) Seed Selection.} It aims to validate \method does not rely on specific seeds.
We select three groups of $<$requirement, code$>$ pairs as seeds and ask an annotator to write SCoTs for them. Then, we obtain three groups of examples. We measure the performance of \method with different groups of examples. \textbf{(2) Writing Style.} People have different writing styles. It aims to validate that \method does not rely on specific writing styles.
We hire three annotators to independently write SCoTs for the same example seed, and obtain three groups of examples. Annotator A is a Ph.D. student in software engineering. Annotator B is a product manager from the industry. Annotator C is a developer from the industry. Then, we measure the performance of \method with different annotators.

For comparison, we also measure the robustness of \cot in the above settings. We select ChatGPT as the base model and conduct evaluations in HumanEval.

\textbf{Results.} 
The results are shown in Table \ref{tab:RQ3_example} and \ref{tab:RQ3_annotator}, respectively.

\textbf{Analyses.}
\uline{\method is robust to examples.} As shown in Table \ref{tab:RQ3_example} and \ref{tab:RQ3_annotator}, \method substantially outperforms \cot when using different example seeds or annotators. It validates that \method does not depend on specific seeds or writing styles. It also shows that the improvements of \method come from the program structures instead of specific details in examples.

We also notice that there are slight variances in the performance of \method under different settings. It is expected for prompting techniques using examples. Similar variances can be found in \method, and \method still outperforms \cot in different settings.

\begin{tcolorbox}[size=title]
\textbf{Answer to RQ3:} \method is robust to examples. Under different example seeds or writing styles, \method substantially outperforms the SOTA baseline - \cot.
\end{tcolorbox}

\subsection{RQ4: What are the contributions of different program
structures in \method?}
\label{sec:results:RQ4}

As stated in Section \ref{sec:approach:SCoT}, \method introduces
basic structures (\ie sequence, branch, and loop) and the input-output structure. This RQ is designed to analyze the contributions of different program structures.

\textbf{Setup.}
We select ChatGPT as the base model. Then, we conduct
an ablation study by independently removing basic structures and the input-output (IO) structure. 
When removing basic structures, we use a CoT with an IO structure as the intermediate steps.
When removing the IO structure, the SCoT only contains a solving process with basic structures.
We select ChatGPT as the base model. 

\textbf{Results.} The results are shown in Table \ref{tab:RQ4}.
``w/o'' is the abbreviation of without.

\begin{figure}[t]
\centering
\includegraphics[width=0.85\linewidth]{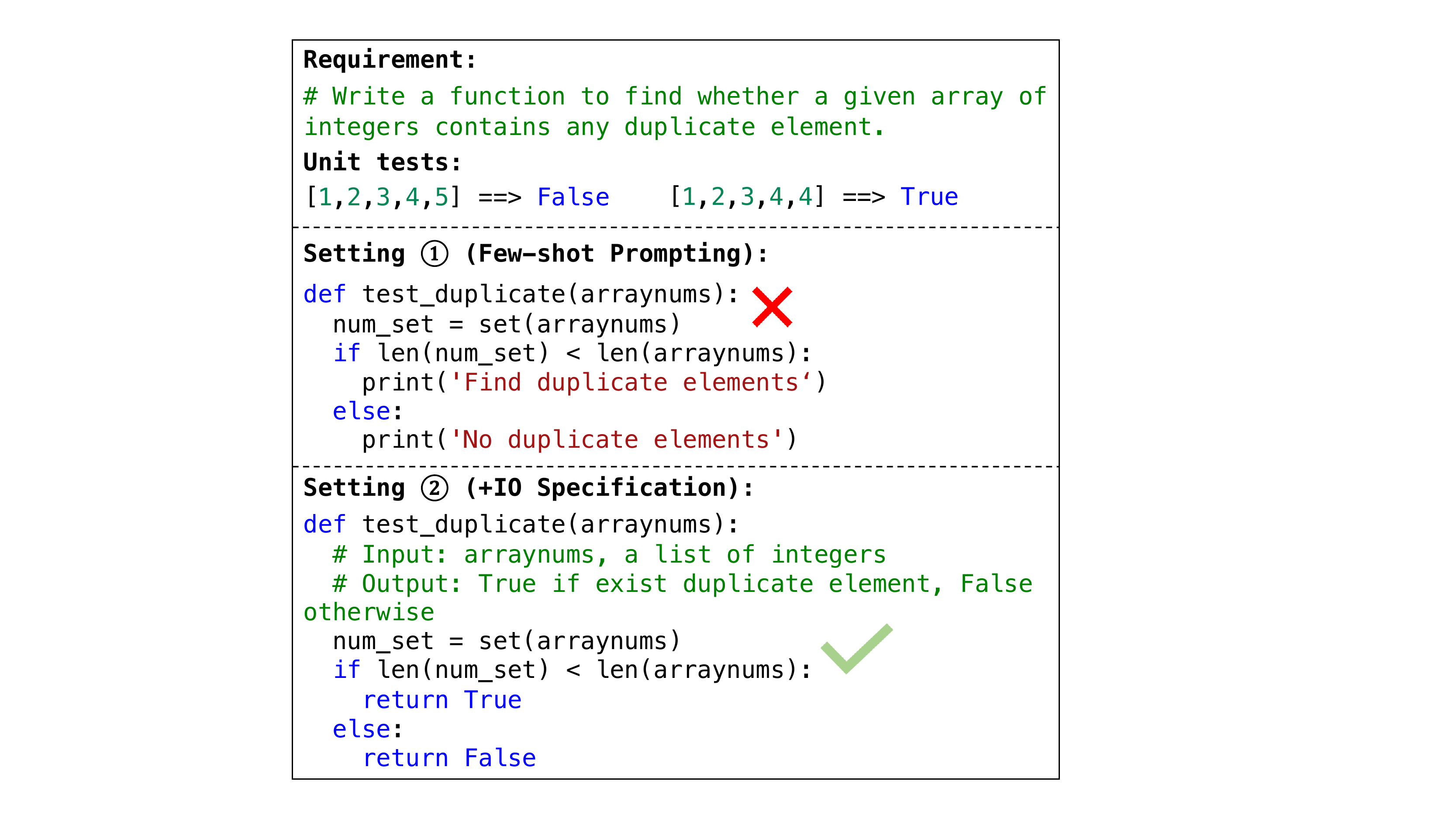}
\vspace{-0.2cm}
\caption{The comparison of \method and \method without basic structures.}
\label{fig:RQ4_ex1}
\end{figure}

\textbf{Analyses.}
\uline{(1) Three basic structures are beneficial to design a feasible solving process.} 
In Table \ref{tab:RQ4}, after removing basic structures, the performance of \method drops obviously. We carefully inspect failed cases and find that LLMs benefit from using basic structures to clearly write a solving process. Figure \ref{fig:RQ4_ex1} shows the intermediate steps of \method and \method without basic structures. \method without basic structures uses CoTs, which sequentially describe how to write the code line by line and contain many ambiguities.
For example, the scopes of two iterations on lines 2 and 4 are unclear. LLMs are likely to misunderstand the CoT and generate incorrect code. In contrast, \method uses three basic structures to describe the solving process. The SCoT is clear and is similar to code, benefiting the following code implementation.

\begin{figure}[t]
\centering
\includegraphics[width=0.85\linewidth]{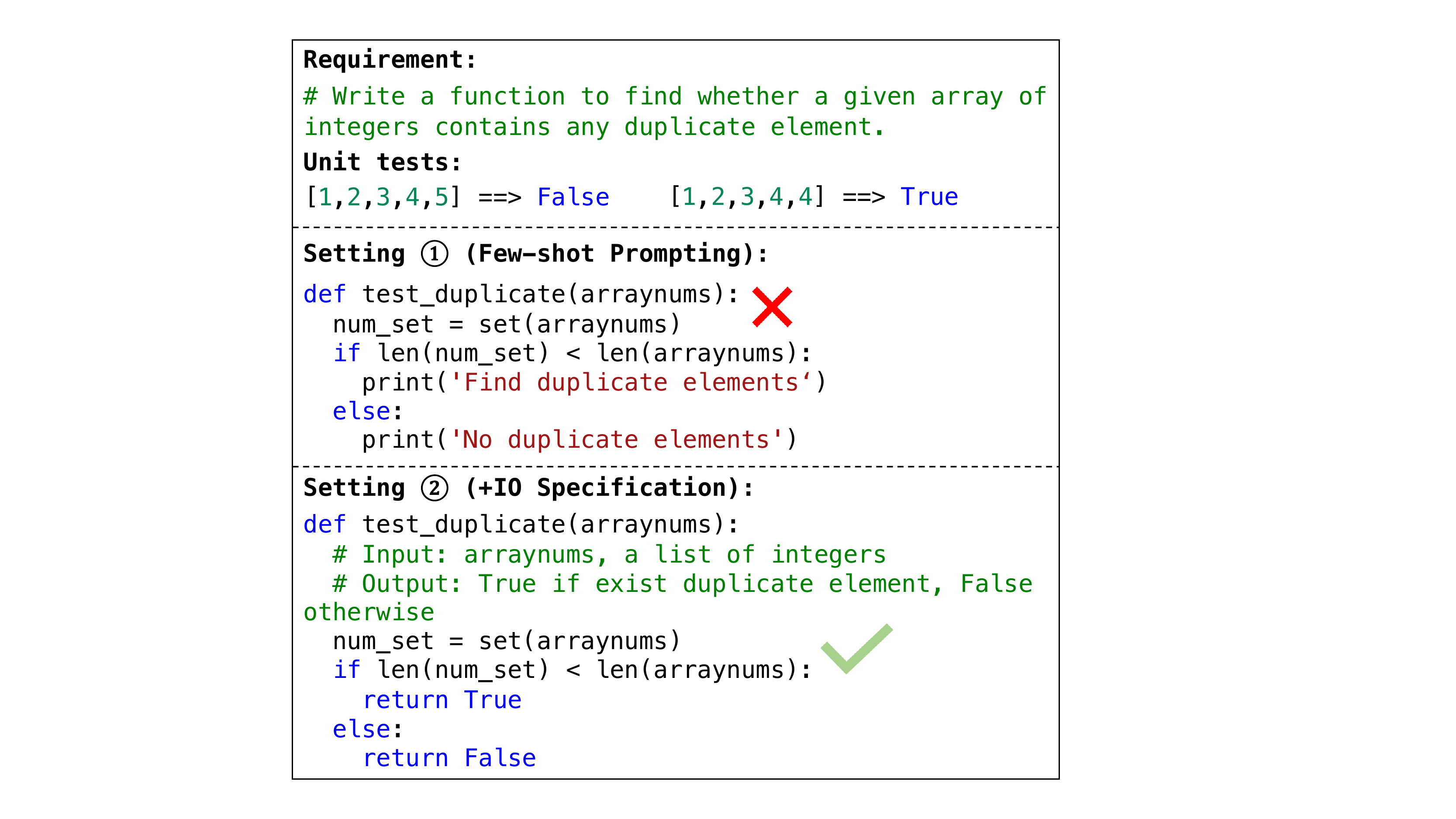}
\vspace{-0.2cm}
\caption{The comparison of \method and \method without the IO structure.}
\label{fig:RQ4_ex2}
\end{figure}

\uline{(2) The IO structure benefits the requirement understanding.}
In Table \ref{tab:RQ4}, after deleting the IO structure, the performance of \method has a slight decrease. We analyze failed cases and think the IO structure benefits the requirement understanding. Figure \ref{fig:RQ4_ex2} shows two programs from \method and \method without the IO structure. We can see that \method without the IO structure wrongly understands the output format and generates an incorrect program. After adding the IO structure, LLMs first reason about the input-output format and correctly return a boolean value.


\begin{table*}[t]
\centering
\caption{The comparison of SCoT-P prompting and \method. The numbers in {\color[HTML]{FE0000} red} denote \method's relative improvements compared to SCoT-P prompting.}
\vspace{-0.3cm}
\begin{tabular}{lccccccccc}
\toprule
\multicolumn{1}{c}{} & \multicolumn{3}{c}{HumanEval} & \multicolumn{3}{c}{MBPP} & \multicolumn{3}{c}{MBCPP} \\
\multicolumn{1}{c}{\multirow{-2}{*}{Approach}} & Pass@1 & Pass@3 & Pass@5 & Pass@1 & Pass@3 & Pass@5 & Pass@1 & Pass@3 & Pass@5 \\
\midrule
\cot & 53.29 & 69.76 & 75.52 & 41.83 & 51.04 & 54.57 & 53.51 & 63.84 & 67.03 \\
SCoT-P prompting & 55.23 & 70.33 & 75.94 & 43.28 & 52.16 & 55.77 & 54.25 & 64.09 & 67.78 \\
\method & \textbf{60.64} & \textbf{73.53} & \textbf{77.32} & \textbf{46.98} & \textbf{55.31} & \textbf{58.36} & \textbf{57.06} & \textbf{65.70} & \textbf{68.70} \\ \midrule
Relative Improvement & {\color[HTML]{FE0000} \textbf{9.80\%}} & {\color[HTML]{FE0000} \textbf{4.55\%}} & {\color[HTML]{FE0000} \textbf{1.82\%}} & {\color[HTML]{FE0000} \textbf{8.55\%}} & {\color[HTML]{FE0000} \textbf{6.04\%}} & {\color[HTML]{FE0000} \textbf{4.64\%}} & {\color[HTML]{FE0000} \textbf{5.18\%}} & {\color[HTML]{FE0000} \textbf{2.51\%}} & {\color[HTML]{FE0000} \textbf{1.36\%}} \\
\bottomrule
\end{tabular}
\vspace{-0.3cm}
\label{tab:PSCoT}
\end{table*}

\begin{tcolorbox}[size=title]
\textbf{Answer to RQ3:} The input-output structure helps LLMs understand requirements and improves ChatGPT by up to 6.37\% in Pass@1. Three basic structures are beneficial to clearly describe a solving process and improve ChatGPT by up to 12.73\% in Pass@1.
\end{tcolorbox}

\vspace{-0.2cm}
\section{Discussion}
\label{sec:discussion}

\subsection{SCoT \textit{vs.} Pseudocode}
\label{sec:discussion:pseudocode}

We notice that the SCoT is similar to the pseudocode. The SCoT and pseudocode both contain an input-output structure and a solving process. We randomly select 100 generated SCoTs and manually review them. We find that 26\% of SCoTs are very close to the pseudocode. On one hand, we think the similarity enhances the usability of our approach. For example, users can quickly know the behavior of a program based on its SCoT. The SCoT also can be inserted into the comment and benefits future maintenance.
On the other hand, the majority of SCoTs (74\%) are different from the pseudocode because they are more abstract. 
Specifically, SCoTs tend to use natural languages to summarize an operation, \eg \texttt{calaluate the sum of list1}.
But the pseudocode contains more implementation details, \eg \texttt{sum $\leftarrow$ 0; for i in list1: sum $\leftarrow$ sum + i;}. 

Compared to the pseudocode, we think the SCoT is a better choice for intermediate steps. Because a SCoT naturally decomposes code generation into two steps. LLMs first focus on exploring diverse solutions and then implement a program in a standardized way. To validate this point, we design a variant of \method, named SCoT-P Prompting. It is similar to \method, but considers the pseudocode as intermediate steps. We apply SCoT-P Prompting and \method to ChatGPT and measure their accuracy.
The results are shown in Table \ref{tab:PSCoT}. \method substantially outperforms SCoT-P Prompting on three benchmarks. The improvements show the superiority of our SCoT.

\vspace{-0.2cm}
\subsection{\method \textit{vs.} Rank Techniques}
\label{sec:discussion:rank}

\begin{figure}[t]
\centering
\includegraphics[width=0.9\linewidth]{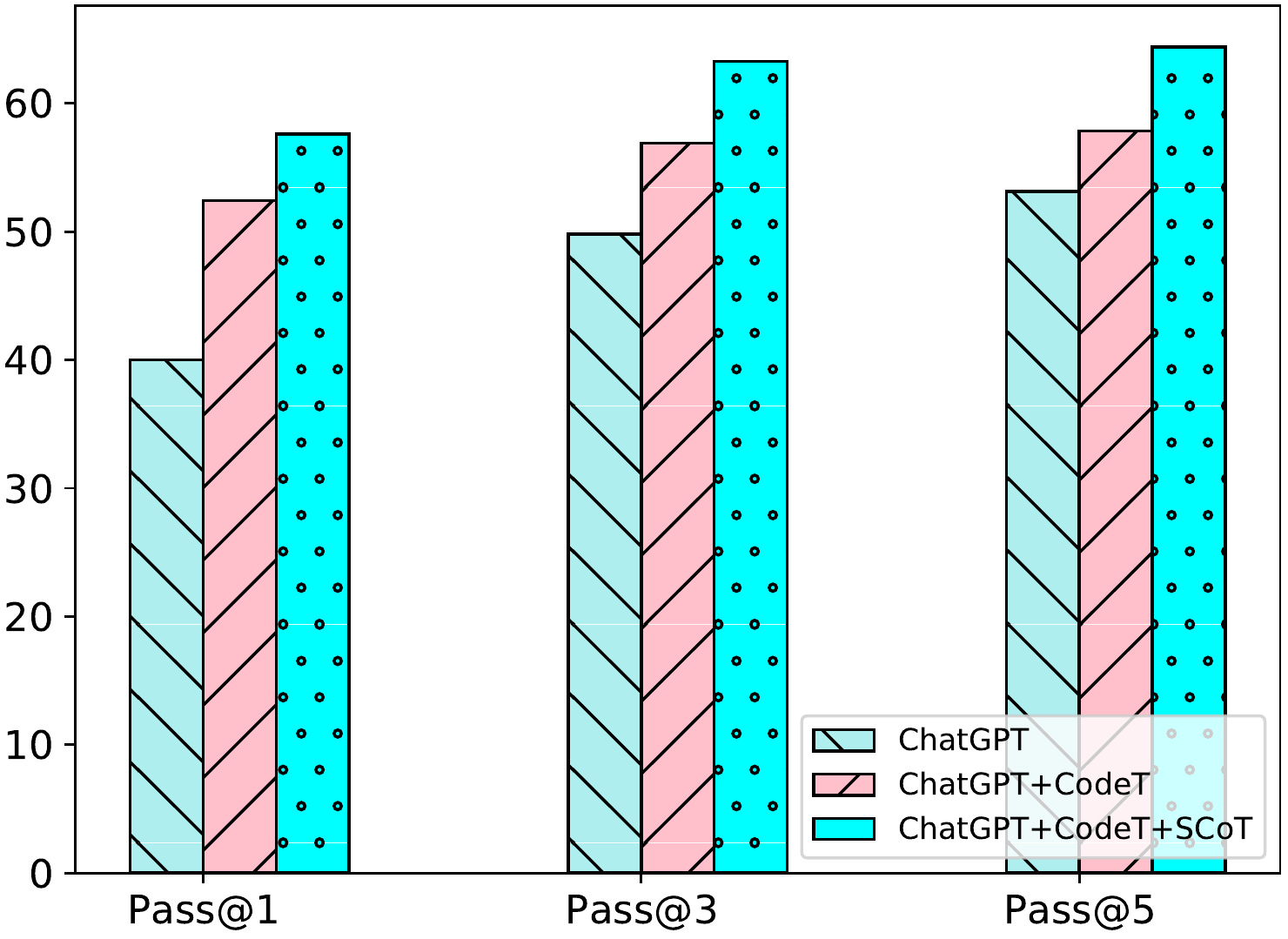}
\vspace{-0.3cm}
\caption{The complementarity between CodeT and \method.}
\label{fig:with_CodeT}
\vspace{-0.2cm}
\end{figure}

Some recent studies \cite{CodeT, fault-aware-ranker} propose \textit{rank techniques} to improve the performance of LLMs on code generation. Given a requirement, they first sample many programs from LLMs and then use test cases or neural networks 
to rerank sampled programs. For example, CodeT \cite{CodeT} is a popular rank technique. CodeT does large-scale sampling and executes sampled programs on auto-generated test cases. Based on execution results, the programs are reranked. In this paper, we do not directly compare our approach to rank techniques due to two reasons.

\uline{(1) \method and rank techniques have different focuses, and they are complementary.}
Our work aims to design a new prompting technique and improve the accuracy of LLMs in code generation. Rank techniques do not care about LLMs and aim to pick the best one from LLMs' multiple outputs. In practice, users can use \method to generate many programs and then use rank techniques to pick a final output. 

To verify the complementarity between \method and rank techniques, we conduct an exploratory experiment. We select ChatGPT as a base model and progressively introduce CodeT and \method. The results on MBPP are shown in Figure \ref{fig:with_CodeT}. We can see that the performance of ChatGPT is continually improved by adding CodeT and \method.

\uline{(2) Rank techniques approaches rely on execution environments.}
Rank techniques require executing programs on test cases and using execution results to rerank programs. In many realistic programming scenarios, users want to get code suggestions for an unfinished project. It is infeasible to execute auto-generated programs. Thus, we think rank techniques have limited application scenarios and make additional use of the execution results. Our approach works in a general scenario and does not use execution results. Thus, it is unfair to directly compare \method to rank techniques.

\vspace{-0.2cm}
\subsection{Threats to Validity}
\label{sec:discussion:threat}

There are three main threats to the validity of our work.

\textbf{(1) The generalizability of experimental results.}
To mitigate this threat, we carefully select the benchmarks, metrics, and baselines. Following previous studies \cite{Codex, MBPP, MBXP}, we pick three representative code generation benchmarks. They are hand-written or collected from real-world programming communities, and cover two popular languages (\ie Python and C++). For evaluation metrics, we select a widely used metric - Pass@$k$, which utilizes test cases to check the correctness of programs. We use the unbiased Pass@$k$ which is more reliable \cite{Codex}. For comparison baselines, we select the SOTA prompting techniques and conduct a comprehensive comparison in Section \ref{sec:results}. \method and baselines have the same example seeds and maximum generation lengths.

\textbf{(2) The impact of the two-step pipeline.} 
\cot generates a CoT and the code in one step. Our \method generates the code in two steps. It first generates SCoTs and then generates the code. It is possible that the improvements come from the two-step pipeline. To solve this threat, we have two considerations. First, LLMs in our experiments are auto-regressive language models. For an auto-regressive language model, a one-step pipeline and a two-step pipeline are theoretically equivalent. Second, we conduct an ablation study in Section \ref{sec:results:RQ4}. We keep the two-step pipeline unchanged and remove program structures. The results in Table \ref{tab:RQ4} show that \method without prompt structures has a significant drop in the Pass@k. It shows that the improvements of \method are brought by program structures instead of the two-step pipeline.

\textbf{(3) The data leakage.} Existing LLMs are trained with extensive code files from open-source communities. It is possible that their training data contains the experimental benchmarks, leading to data leakage. But we think that it does not affect the fairness of our experiments. In this paper, we select a specific LLM (\eg ChatGPT) as the base model and apply different prompting techniques to it. Thus, the reported relative improvements between baselines and our approach are credible. In the future, we will add the latest benchmarks to alleviate this threat.

\vspace{-0.2cm}
\section{Related Work}
\label{sec:related}

\textbf{Large language models (LLMs) for Source Code} are large-scale neural networks that are pre-trained with a large corpus consisting of natural language text and source code. Nowadays, LLMs for source code have been expanding and can be divided into two categories: standard language models and instruction-tuned models.

\textbf{Standard Language models} are pre-trained on a large-scale corpus with the next-token prediction objective. They are mainly used to continually complete the given context, such as code completion.
After the success of GPT series \cite{GPT, GPT-2, GPT-3} in NLP, 
OpenAI fine-tunes GPT models on code to produce closed-source Codex \cite{Codex}.
There follow many open-source replication attempts, \eg CodeParrot \cite{CodeParrot}, CodeGen \cite{CodeGen}, CodeGeeX \cite{CodeGeeX}, InCoder \cite{InCoder}, StarCoder \cite{StarCoder} and CodeT5+ \cite{CodeT5+}.

\textbf{Instruction-tuned models} are models after instruction tuning \cite{instruct-tuning}. Instruction tuning trains models to understand human users’ instructions and perform tasks by following instructions.
ChatGPT \cite{ChatGPT} is trained with human feedback \cite{RLHF}, powerful on both natural language tasks and programming tasks.
Many attempt to train an ``open-source ChatGPT''.
Alpaca \cite{alpaca} is LLaMA \cite{LLaMA} tuned using self-instruct \cite{self-instruct} and ChatGPT feedback.
Code Alpaca \cite{code-alpaca} is LLaMA tuned using self-instruct and ChatGPT feedback, with instructions focusing on programming tasks.
WizardCoder \cite{WizardCoder} is StarCoder \cite{StarCoder} tuned using Evol-Instruct \cite{WizardLM} and ChatGPT feedback with Code Alpaca's dataset as seed dataset.
InstructCodeT5+ \cite{CodeT5+} is CodeT5+ \cite{CodeT5+} tuned on Code Alpaca's dataset.




\textbf{Prompting Techniques.}
With the enormous number of parameters (\eg Codex: 175 billion parameters), it is hard to directly fine-tune LLMs on code generation. \textit{Prompting techniques} are a popular approach, which leverages LLMs to generate code by inputting a special prompt.

Early, researchers proposed zero-shot prompting and few-shot prompting. Zero-shot prompting concatenates a task instruction (\eg \texttt{please generate a program based on the requirement}) and a requirement together, making a prompt. Based on the zero-shot prompting, few-shot prompting further adds several $\left<\right.$requirement, code$\left.\right>$ pairs to the prompts, so that LLMs can learn code generation from given examples.



The Chain-of-Thought (CoT) prompting \cite{chain-of-thought} is a recently proposed prompting technique. 
\cot asks LLMs first to generate CoTs (\ie intermediate natural language reasoning steps) and then output the final code.
It allows LLMs to first design a solving process that leads to the code. \cot has achieved the SOTA results in natural language generation and sparked lots of follow-up research, such as self-consistency prompting \cite{Self-Consistency}, least-to-most prompting \cite{Least-to-Most}. But these prompting techniques are designed for natural language generation and bring slight improvements in code generation.

In this paper, we propose a novel prompting technique named Structured Chain-of-Thought (SCoT) prompting. Different from standard \cot, \method explicitly introduces program structures and asks LLMs to generate intermediate reasoning steps with program structures. We compare \cot and \method in Section \ref{sec:results}. The results show that \method significantly outperforms \cot in three benchmarks.

\vspace{-0.2cm}
\section{Conclusion and Future Work}
\label{sec:conclusion}

Large Language Models (LLMs) with Chain-of-Thought (CoT) prompting is the state-of-the-art (SOTA) approach to generating code. It first generates a CoT and then outputs the code. A CoT is several intermediate natural language reasoning steps. However, CoT prompting still has low accuracy in code generation. This paper proposes a Structured CoT (SCoT) and presents a new prompting technique for code generation, named \method. \method asks LLMs to generate a SCoT using program structures (\ie sequence, branch, and loop structures). Then, LLMs generate the code based on the SCoT.
A large-scale study on three benchmarks shows that \method significantly outperforms \cot in Pass@k and human evaluation. Besides, \method is robust to examples and obtains stable improvements.

In the future, we will explore new prompting techniques for code generation. For example, source code can be represented by a tree (\eg abstract syntax tree). We can design a tree-based prompting technique, which uses LLMs to generate a tree.


\normalem
\bibliographystyle{ACM-Reference-Format}
\bibliography{reference}

\appendix

\end{document}